\documentclass[]{article}
\usepackage{graphicx}                    % For eps figures, newer & more powerfull
\usepackage[square]{natbib}
\begin{document}
\title{A Study of Halo Coronal Mass Ejections and Related Flare and Radio Burst Observations in Solar Cycle 23}

\author{M. Georgiou, E. Mitsakou, G. Pothitakis, \\A. Hillaris, P. Preka-Papadema, X. Moussas\\
\emph{University of Athens, 15784, Panepistimiopolis Zografos, Athens, Greece}}
\maketitle
\begin{abstract}
We present a statistical study of dynamical and kinetic characteristics of CMEs which show temporal and spatial association with flares and type II radio bursts or complex radio events of type II bursts and type IV continua. This study is based on a set of earth-directed full halo CMEs occurring during the present solar cycle, with data from the Large Angle Spectrometric Coronagraphs (LASCO) and Extreme-Ultraviolet Imaging Telescope (EIT) aboard the Solar and Heliospheric Observatory (SOHO) mission and the Magnetic Fields Investigation (MFI) and 3-D Plasma and Energetic Particle Analyzer Investigation experiment on board the WIND spacecraft.
\end{abstract}

\section{Introduction}
Coronal Mass Ejections (CMEs) are solar transients characterised by the expulsion of large quantities of plasma and frozen in magnetic field from the corona to the interplanetary space. The response of the heliosphere to the CMEs is manifested as magnetic clouds, IP shocks and energetic particle events (SEPs), which in turn, constitute the {\emph{Space Weather}} 
(cf. for example \cite{Messerotti}, \cite{Lathuillere}). 
The influence on {\emph{Space Weather}} of Earth--Bound halo CMEs  is significant, as they are moving along the Sun--Earth line; meanwhile, these constitute only a small fraction {$\approx 3,5\%$} of all CMEs observed. Solar radio bursts, on the other hand, represent an efficient diagnostic of energetic phenomena on the Sun. Type II bursts result either from a flare blast wave or are driven by a CME (\cite{Classen}), while the type IV continua originate from energetic 
electrons trapped within plasmoids, magnetic structures or substructures of CMEs (\cite{Bastian}). 

In this paper we present a statistical analysis with emphasis on the correlation between the earth-directed halo CMEs, metric type II and type IV radio bursts and flares. 

\section{Data Analysis \& Results}

The occurence of a halo CME in the time interval from January 1996 to December 2003 was identified from the LASCO catalogues online (\cite{Yashiro}); the CMEs parameters,which included the CME's linear speed, acceleration and take off time, were obtained from the same catalogues. Secondly, we used five minutes averaged data of the MFI and 3-D Plasma and Energetic Particle Analyzer Investigation experiment onboard the WIND spacecraft to identify the passage of the internal material of the CME through the following distinctive signatures: a transient shock marked by an abrupt increase in the strength of the interplanetary magnetic field, which was complemented by a depression of the density, temperature and speed of the particles populations within the post-shock flow. As a result, 67 earth-directed full halo CMEs were identified for this study.
%-------------------------------------------------------------------------------------
\begin{figure}[h!]
\begin{minipage}[t]{6cm}
\begin{center}
\includegraphics[width=6cm]{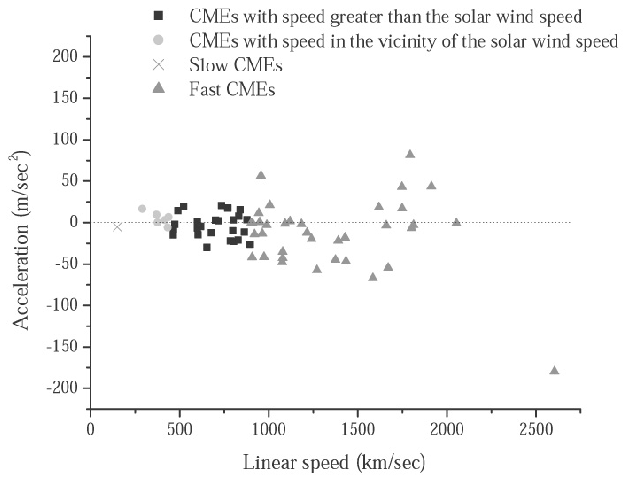}
\end{center}
\end{minipage}
\hfill
\begin{minipage}[t]{6cm}
\begin{center}
\includegraphics[width=6cm]{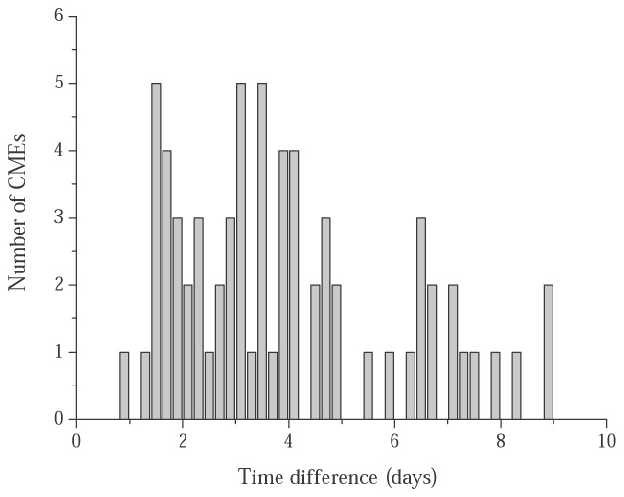}
\end{center}
\end{minipage}
\begin{minipage}[t]{6cm}
\begin{center}
\includegraphics[width=6cm]{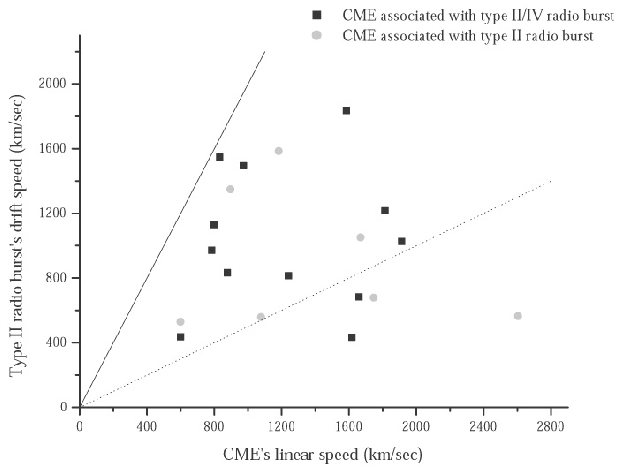}
\end{center}
\end{minipage}
\hfill
\begin{minipage}[t]{6cm}
\begin{center}
\includegraphics[width=6cm]{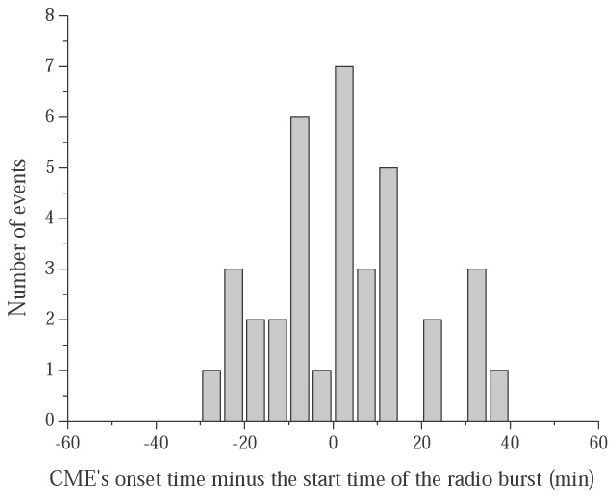}
\end{center}
\end{minipage}
  \caption{\textit{Upper left panel}: CME acceleration as a function of the CME's linear speed. \textit{Upper right panel}:Histogram of time elapsed between the observation of the CME by LASCO coronagraphs and the detection of the material expelled by WIND instruments. The time difference ranged from less than one day to more than four days. \textit{Lower left panel}: Type II radio burst's drift speed versus the CME's linear speed. 
 \textit{Lower right panel}: Histogram (with bin size of 5 minutes) of time difference between the time when the CME was at 1 Ro and the radio burst's start time.\label{CME_Velocity}}
 \end{figure}
%-------------------------------------------------------------------------------------

The movies of EUV images obtained by the SOHO/EIT, Yohkoh SXT images and Ha filtergrams were examined for associated activity on the visible solar disk within a time window of 60 minutes before and after the extrapolated CME take off time. The associated solar flares properties were derived using data from the Soft X-rays Sensor (SXR) aboard the GOES satellite. All of the associated flares were simple events without multiple peaks, or truncated by the occurrence of another flare. The duration of each flare was measured using the GOES 1-8A channel and it was defined as the time between the point were the gradient of the X-rays flux turned positive, to the point where it returned to the pre--flare value or leveled off. The peak colour temperature for each flare event was estimated from the ratio of fluxes in the two GOES channels,  following the standard calibration of {\emph{Garcia}} \cite{Garcia94}. In this study, we focused our attention on two forms of activity, the flares and the metric type II and IV radio bursts. A 60 minutes time window between the CME's extrapolated take off time and the radio event's onset time was introduced as the criterion for establishing associations. Ground--based observations of the II and IV metric radio bursts recorded in SGD, were used for the estimation of the radio bursts parameters. The type II frequency drift and, in turn, the radial speed, was derived using a density model of the solar corona by {\emph{Newkirk}} (\cite{Newkirk67} and  \cite{Newkirk61}).
%--------------------------------------------------------------------------------------
\begin{figure}[h!]
\begin{minipage}[t]{6cm}
\begin{center}
\includegraphics[width=6cm]{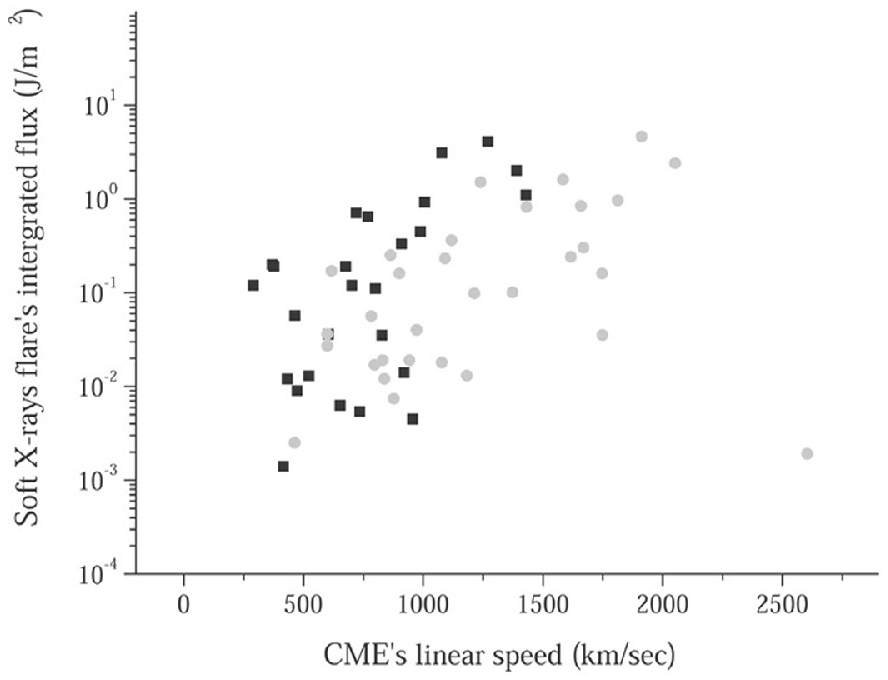}
\end{center}
\end{minipage}
\hfill
\begin{minipage}[t]{6cm}
\begin{center}
\includegraphics[width=6cm]{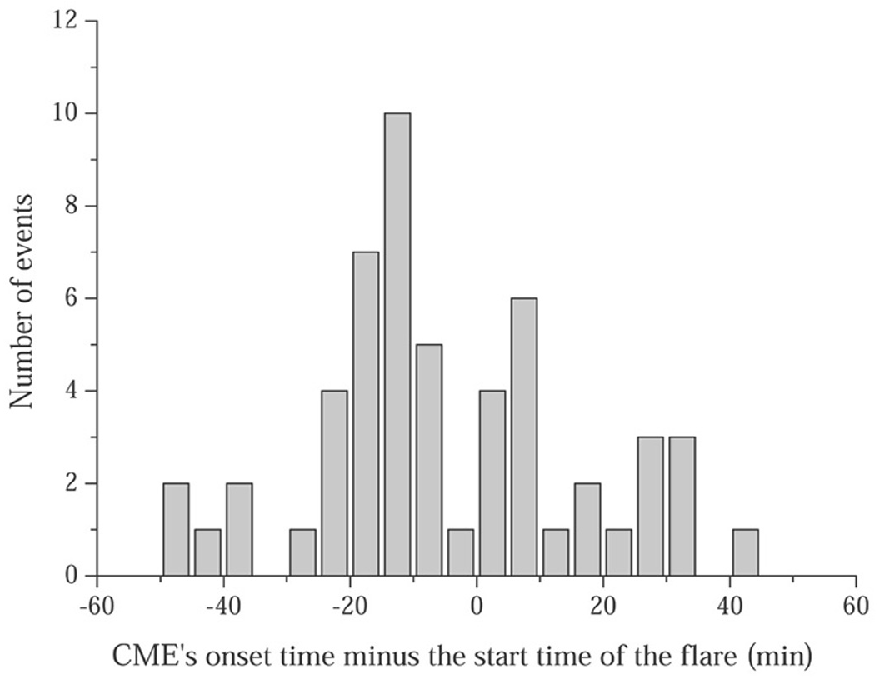}
\end{center}
\end{minipage}
\begin{minipage}[t]{6cm}
\begin{center}
\includegraphics[width=6cm]{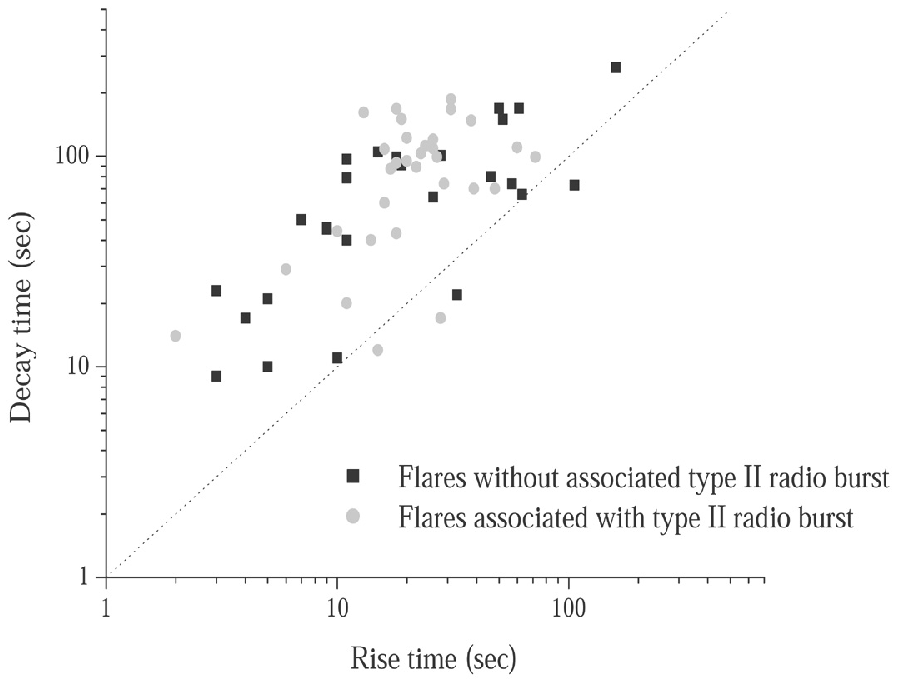}
\end{center}
\end{minipage}
\hfill
\begin{minipage}[t]{6cm}
\begin{center}
\includegraphics[width=6cm]{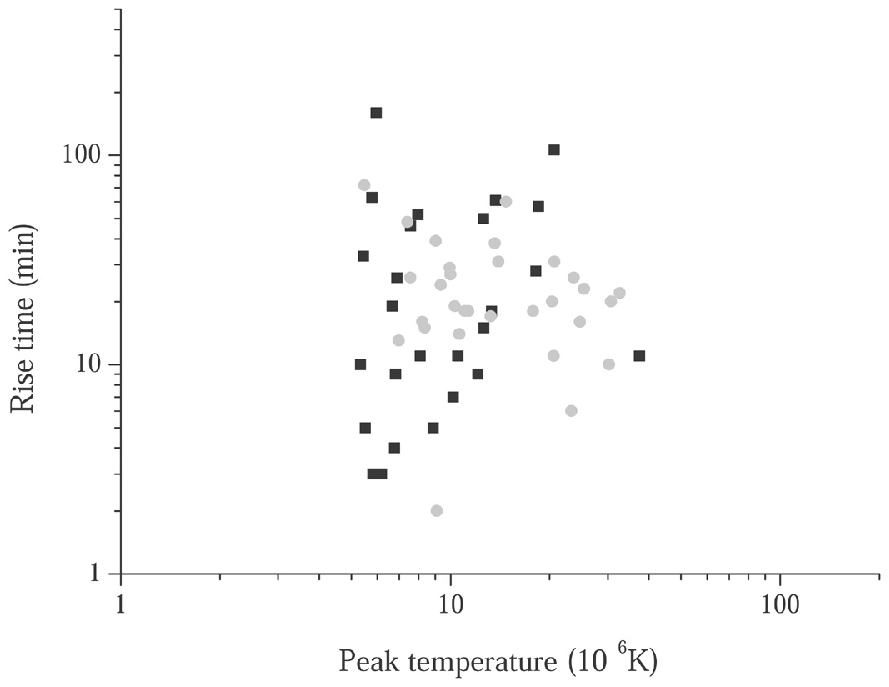}
\end{center}
\end{minipage}
\caption{\textit{Upper right panel}: Histogram (with bin size 5 minutes) of the time elapsed between the onset 
of the CME and the start time of the soft X-rays flare. 
\textit{Upper left panel}: Soft X-rays flare's intergrated flux versus the CME's linear speed. 
 \textit{Lower left panel}: The time from the start of the flare to the time of flare's peak 
intensity versus the length of the the subsequent decrease in the flare's intensity.
\textit{Lower right panel}: The length of the initial increase in the flare's intensity versus the flare's peak temperature.}
\label{CME_FLARE}
\end{figure}
%-------------------------------------------------------------------------------------

The CME's linear speed provides an estimation of the mean speed within the coronagraph's field of view and in our list it ranged from $150~km/s$ to $2600~km/s$. With an average value of $\approx1004~km/s$, twice the average speed ($\approx489~km/s$) of all the CMEs observed in the same period (\cite{Gopal03}), the CMEs in our list constitute a fraction of faster, hence, more energetic than average CMEs. CMEs with speed greater than the typically slow solar wind ($450~km/s < Vcme < 900~km/s$) show predominant deceleration ($a\approx-3.944m/s^2$), as well as, fast CMEs ($900 km/s < Vcme$) with an average deceleration of $a\approx-13.487m/s^2$, indicating that the retarding exceed the propelling forces than drive the CME outward in the heliosphere. On the other hand slower CMEs with speed in the vicinity of the solar wind speed ( $250 km/s < Vcme < 450 km/s$) were found to have small acceleration ($a\approx4.87~m/s^2$)(cf. figure \ref{CME_Velocity}, left panel).

Observations of metric radio bursts were compared with the observations of the halo CMEs and 42 of these events ($\approx63\%$) showed a close relation to type II radio bursts with respect to speed and extrapolated onset time (cf.figure \ref{CME_Velocity}). Type II frequency drift rates appeared to be all in excess of $400~km/s$ and the mean speed was found to be equal to $\approx1009~km/s$ . Furthermore, a CME related group of 32 events ($\approx48\%$) were accompanied by a type IV continuum. Considering the time difference between the time when the CME was at 1 Ro and the radio burst's start time, 27 of CMEs take offs preceded and 19 followed their corresponding radio burst onset; the mean value of the time difference was found within 0 to 5 minutes from the CMEs take off .
%--------------------------------------------------------------------------------------
\begin{figure}[h!]
\begin{minipage}[t]{6cm}
\begin{center}
\includegraphics[width=6cm]{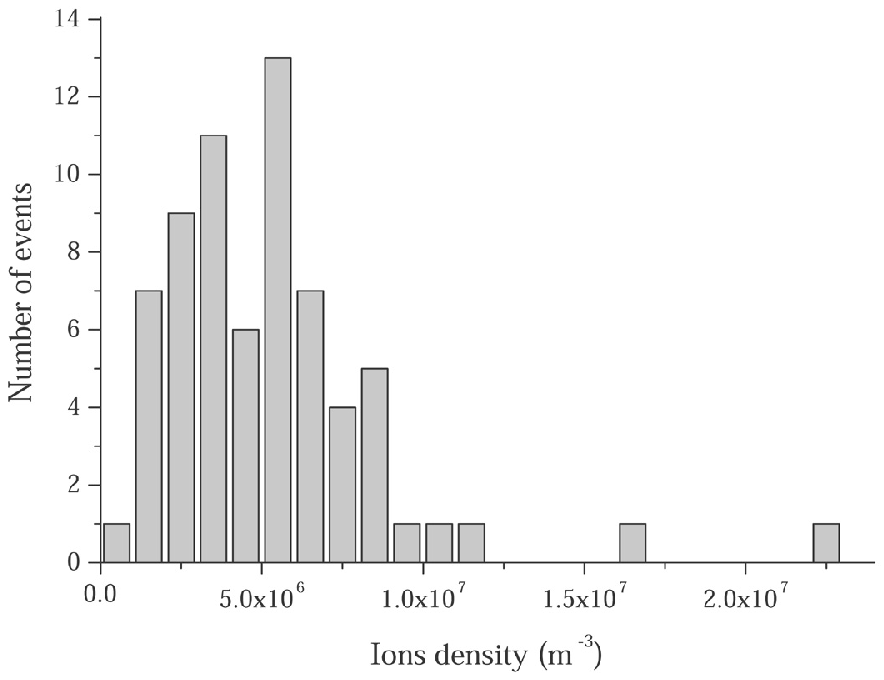}
\end{center}
\end{minipage}
\hfill
\begin{minipage}[t]{6cm}
\begin{center}
\includegraphics[width=6cm]{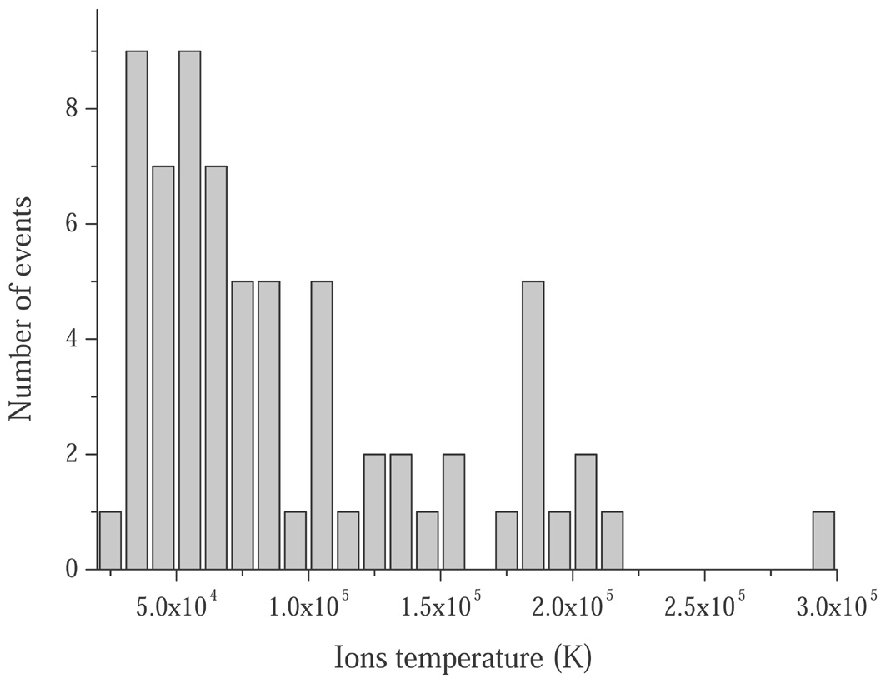}
\end{center}
\end{minipage}
\begin{minipage}[t]{6cm}
\begin{center}
\includegraphics[width=6cm]{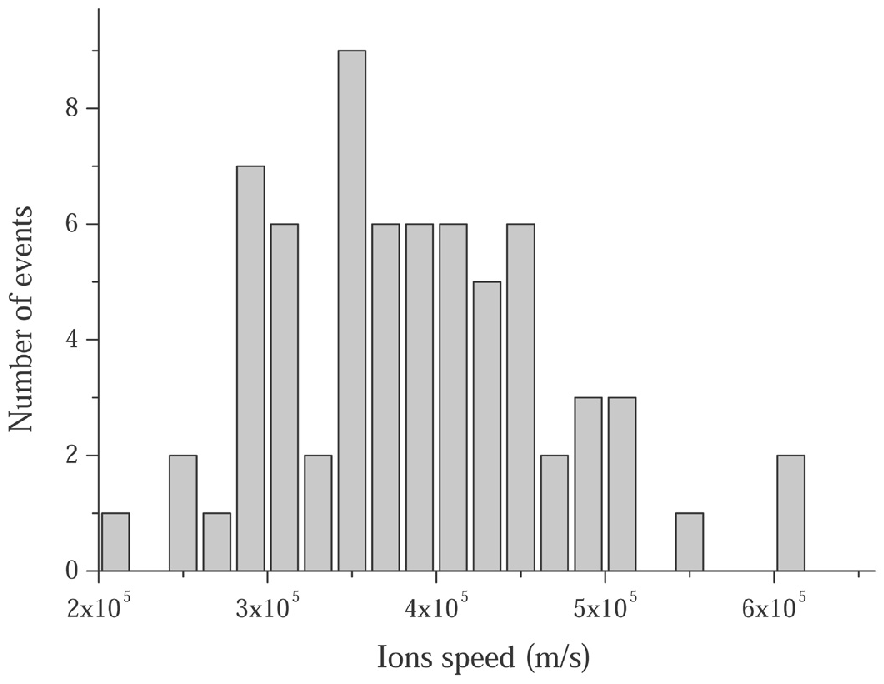}
\end{center}
\end{minipage}
\hfill
\begin{minipage}[t]{6cm}
\begin{center}
\includegraphics[width=6cm]{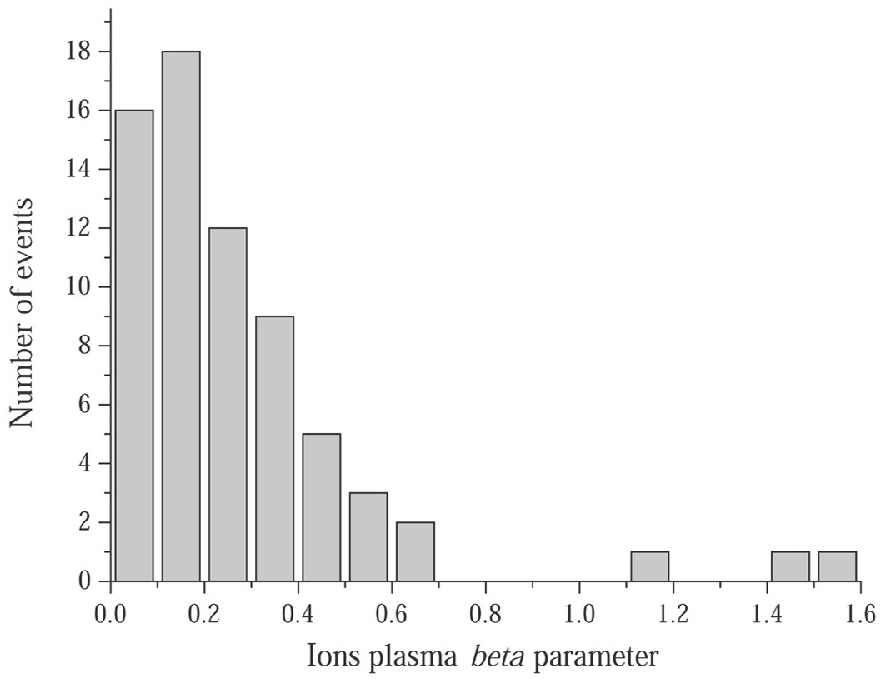}
\end{center}
\end{minipage}
  \caption{The average plasma properties within the CMEs observed by the WIND. From top left to bottom right are the distribution of the ions mean density, the ions mean temperature, the average speed and the average {\emph{beta}} parameter within the ions population of the CMEs.}
\label{ICME parameters}
\end{figure}
%--------------------------------------------------------------------------------------------

The majority of the sampled CMEs were associated with surface activity. In this study, the Ha and soft X--rays flares were the only form of activity taken into account.Flares of all sizes (the X-rays importance ranged from B to X), occurring at all longitudes (19 flares originated within 25 degrees from the central meridian and 5 flares within 30 degrees from the solar limb) were found to be associated with the CMEs. 57 of the CMEs ($\approx85\%$) were associated with soft X-rays flares and 28 ($\approx41\%$) were associated with a flare and a type II radio burst. 45 of the CMEs were associated with Ha flares ($\approx67\%$). For all of the events appears a general correlation between the CME's linear speed and the soft X-rays integrated flux. The events associated with type II radio burst are clustered toward the greater velocity and flux ranges, while the CME speed exceeds a threshold of about 500km/s. A comparison between the relative times of the initial increase and the subsequent decrease in the flare's intensity indicates a clear corellation, meanwhile it appears to be a systematic difference between the two times, with the decrease lasting longer than the initial increase. 18 flares lasted longer than 180 minutes, while 18 events as short as 60 minutes and less also show to be associated with CMEs. As regards the peak temperature of the flares, events without associated radio emission tend to reach peak temperature faster than those associated with a radio burst. Considering the time elapsed between the onset of the CME and the start time of the flare, 23 of the CMEs lift offs preceded and 34 followed their corresponding flare onset. The mean of the distribution was found within 10 to 5 minutes before the CMEs take off (cf. figure \ref{CME_FLARE}).

The size of the interplanetary CMEs in the radial direction observed near the Earth's orbit ranged from 0.078A.U. to 0.58A.U. with a mean value of 0.26A.U., while the size of the core of the interplanetary post-shock flow ranged from 0.055A.U. to 0.36A.U. with a mean value of 0.17A.U.. The enormous radial size of the interplanetary CMEs cannot be considered to be typical of those observed close to the Sun, indicating an expansion of the CME in the heliosphere, which causes their radial size to increase with distance and may affect the average physical conditions within the CMEs.

The average density of ions inside the CMEs is about $5.3\times10^6 ions/cm^3$, slightly lower than the average background solar wind density, $9\times10^6 ions/cm^3$, which is in accordance with the CMEs expansion. The average density of ions ranged from $0.93\times10^6 ions/cm^3$ to $2.2\times10^7 ions/cm^3$. The average value of the ions temperature within the CMEs was estimated of about $10^6 K$, slightly lower than the average solar wind temperature, $1.2 10^6 K$, which is in accordance with the low temperature criterion used for the identification of the CMEs in the interplanetary medium. The average ions temperature ranged from $2\times10^4 K$ to $3\times10^5 K$. The average speed within the ions population of the CMEs lies in the range from $220~km/s$ to $620~km/s$ with a symmetrical distribution around their mean value of $Vi = 384 km/s$. Most CMEs were found to have a propagation speed similar to that of the average solar wind speed in the same period, $Vsw = 340 km/s$, and only a few had speed above the speed of the slow flow of the solar wind (cf. figure \ref{ICME parameters}). These results are similar to the calculations of CMEs observed by WIND and ACE in the same time interval by {\emph{Liu et al.}}(\cite{Liu}) and the results obtained from Helios 1 and 2 in the inner heliosphere(\cite{Bothner}). The interplanetary post-shock flow is characterised by low plasma {\emph{beta}} values between 0.025 and 1.6 with a mean value of 0.28 , as these were calculated from the ions parameters only. The plasma's magnetic pressure dominated the thermal pressure within the CMEs of our data set, due to their higher than average solar wind magnetic field strength and their unusually low plasma temperature and density, thus, contriduting to the radial expansion of the CMEs.

\section{Conclusions}
We have studied a fraction of full halo CMEs with distinctive multiple signature in the interplanetary medium, which constitute a population of fast and wide CMEs. The CMEs associated with coronal type II radio bursts are faster than the CMEs without any associated metric radio emission (cf. figure \ref{CME_FLARE}). The average speed of the CMEs associated with type II radio bursts was estimated of about $1200~km/s$, greater than the average speed of the CMEs without associated radio emission ($\approx810~km/s$). Depending on the CME parameters and the interplanetary medium, a shock can be formed in the inner corona, which produces metric type II bursts. It must be noted that almost all of the CMEs with associated type II radio emission (28 out of 32 events) were also associated with an X-rays flare. The close coincidence and other correlations between flares, CMEs and type II radio bursts can be considered to suggest that the influence of flares in the production of type II bursts cannot be ruled out.

\end{document}